\documentclass[%
 aip,
  pop,
 reprint,%
]{revtex4-1}
\usepackage[
  pdfusetitle,
  unicode=true,
]{hyperref}
\usepackage{amsmath,amssymb}
\usepackage{graphicx}
\usepackage{dcolumn}
\usepackage{bm}
\usepackage{etoolbox}

\usepackage{color}
\def \bi #1{{\bm #1}} 
\def \rmd{\mathrm{d}}

\def \ad{\mathrm{ad}}

\makeatletter
\def\@email#1#2{%
 \endgroup
 \patchcmd{\titleblock@produce}
  {\frontmatter@RRAPformat}
  {\frontmatter@RRAPformat{\produce@RRAP{*#1\href{mailto:#2}{#2}}}\frontmatter@RRAPformat}
  {}{}
}%
\makeatother
\begin{document}

\preprint{AIP/123-QED}

\title[Hierarchy of Casimirs]{Hierarchical foliation of one-dimensional Vlasov-{Poisson} system}
\author{K. Maekaku}
\email{maekaku.koki20@ae.k.u-tokyo.ac.jp}
 \affiliation{Graduate School of Frontier Sciences, University of Tokyo, Kashiwa, Chiba 277-8561, Japan}
 
\author{Z. Yoshida}%
\affiliation{ 
National Institute for Fusion Science, Toki 509-5292, Japan
}%

\date{\today}

\begin{abstract}
We elucidate the intermediate of the macroscopic fluid model and the microscopic kinetic model by studying the Poisson algebraic structure of the one-dimensional Vlasov--Poisson system. The water-bag model helps  \textcolor{black}{formulating} the hierarchy of sub-algebras\textcolor{black}{, which interpolates} the gap between the fluid and kinetic models. By analyzing \textcolor{black}{the embedding of} the sub-manifold of an intermediate hierarchy in a more microscopic hierarchy, we characterize the microscopic effect as the symmetry breaking pertinent to a macroscopic invariant.
\end{abstract}

\maketitle

\section{Introduction}

\textcolor{black}{In this study, we} elucidate the hierarchical structure interpolating the macroscopic fluid model and the microscopic Vlasov model of collisionless plasmas\textcolor{black}{. We further} describe kinetic effects as a symmetry\textcolor{black}{-}breaking \textcolor{black}{event, which converts} a larger-scale hierarchy to a smaller-scale one. The \textcolor{black}{core} idea \textcolor{black}{of out approach here} is to formulate a series of self-consistent subsystems (sub-algebras) of the Vlasov Lie--Poisson algebra, and to characterize each subsystem by Casimir invariants\textcolor{black}{. As such, the} conservation of a Casimir invariant \textcolor{black}{would reflect} a particular symmetry in the distribution function\textcolor{black}{. Accordingly,} the kinetic effect breaking such a symmetry \textcolor{black}{would thus }manifests as \textcolor{black}{a} non-conservation of the corresponding Casimir invariant.

Conventionally, the \textcolor{black}{relationship} between the fluid and kinetic models is discussed by invoking the velocity-space moments of the distribution function $f$ (with non-negative exponents $s_1$, $s_2$, and $s_3$)
\[
P_{s_1 s_2 s_3}=\int v_1^{s_1}v_2^{s_2}v_3^{s_3} f(\bi x,\bi v,t)\rmd^3 v ,
\]
and imposing a \emph{closure relation}. The lower-order moments constitute the fluid-mechanical variables (\textcolor{black}{i.e.} density, fluid velocity, pressure tensor), while the higher-order moments measure some deviation of $f$ from Gaussian\textcolor{black}{. It follows that} the moment hierarchy is useful for quantifying the probability distribution caused by collisions. A modern approach considers the Hamiltonian closure.\cite{tassi2017hamiltonian,perin2014closure,perin2015hamiltonian}

\textcolor{black}{Nonetheless, the problem is considered in this study} from the perspective of the Poisson algebra\textcolor{black}{, which} governs the collisionless dynamics of plasmas. \textcolor{black}{Rather than using} the moment hierarchy, we construct a hierarchy of sub-algebras of the Poisson algebra, with the lowest-dimension system corresponding to the fluid model. \textcolor{black}{In practice}, we consider the reduction of the Vlasov Lie--Poisson system by restricting the distribution function $f$ to \textcolor{black}{the} sum of step functions in the velocity space (to be denoted by $V$). \textcolor{black}{According} to Liouville's theorem, the height of each flat top of $f$ remains constant \textcolor{black}{, whereas} the area of the plateau in the $V$-space may change. Hence, such a system of distribution functions defines a sub-algebra (closed subsystem) of the Vlasov system\cite{chandre2013hamiltonian,morrison1980maxwell} 
(see Refs.~\citenum{marsden1974reduction,morrison1998hamiltonian,morrison1982poisson} 
for the general idea of reduction of a Hamiltonian system). \textcolor{black}{Interestingly, a single-plateau system corresponds to the ``fluid model’’\textcolor{black}{, which assumes} only one velocity at each point of the configuration space.} \textcolor{black}{A} hierarchy of sub-algebras \textcolor{black}{is formulated by increasing the number of plateaus.} \textcolor{black}{The limit of an infinite number of plateaus recovers the Vlasov system (in the sense of a finite-difference approximation of the distribution function with respect to the velocity coordinate).} Each subsystem is a leaf of the Vlasov system, which is characterized by Casimir invariants. Generally, a Casimir invariant is a generator of some gauge symmetry pertinent to reduction;\cite{topological,yoshida2021helicity,tanehashi2015gauge}\textcolor{black}{. When} used as a Hamiltonian in a higher-order system, the corresponding Hamiltonian flow keeps the reduced variables constant. In other words, the breaking of the gauge symmetry and the variance of the Casimir \textcolor{black}{invariant} imply the higher-order effect\textcolor{black}{, which} violates the closedness of the lower-order (macroscopic) subsystem.

As a simple and analytically tractable example, we study the water-bag model of a one-dimensional charged-beam system.\cite{wb1,wb2,WB,tennyson1994waterbag} A series of water-bag models with different numbers of water\textcolor{black}{-}bags \textcolor{black}{define} the hierarchical Vlasov sub-algebra.


\section{Preliminaries}
\label{sec:preliminary}

\subsection{Hamiltonian mechanics and Poisson manifold}

In Hamiltonian mechanics, the equation of motion is \textcolor{black}{given by}
\begin{eqnarray}
\dot{\bi{z}} = J \partial_{\bi{z}}H(\bi{z}),
\end{eqnarray}
where $\bi{z} \in M$ is the state vector\textcolor{black}{, whose} totality $M$ is the phase space, $J$ is the Poisson operator, $H$ is the Hamiltonian, and $\partial_{\bi{z}} H$ is the gradient of $H$. We define the Poisson bracket \textcolor{black}{as}
\begin{eqnarray}
[ G, H ] = \langle \partial_{\bi{z}} G, J \partial_{\bi{z}} H \rangle ,
\label{Poisson_bracket}
\end{eqnarray}
and the Poisson operator $J$ must be defined appropriately for this bracket to define a Poisson algebra (\textcolor{black}{i.e.} Lie algebra with Leibniz property). Endowing the function space $C^\infty(M)$ (a member $G(\bi{z})\in C^\infty(M)$ represents a physical quantity and is called an observable) with the Poisson bracket, we call $M$ the Poisson manifold.

In the \emph{canonical} Hamiltonian system, $\bi{z}$ is given as a conjugate $\bi{z}=(\bi{q},\bi{p})^{\mathrm{T}}$, and
\begin{eqnarray}
J= J_c = \left(\begin{array}{cc}0 & I \\-I&0\end{array}\right)\mbox.
\end{eqnarray}
However, there are many possible \emph{non-canonical} Hamiltonian systems (or degenerate Poisson algebras)\textcolor{black}{, which} are defined by more-complicated $J$ \textcolor{black}{expressions. In} general, $J$ may depend on $\bi{z}$ and may have nontrivial kernels (i.e., $\mathrm{rank}\,J < \mathrm{dim}\,M$). As we will show by an example (Sec.~\ref{subsec:reduction}), such a \emph{degeneracy} is often \textcolor{black}{included} by some \emph{reduction} from a higher-dimensional canonical system\textcolor{black}{. Here,} the reduction \textcolor{black}{refers to admitting} observables with only a restricted dependence on $\bi{z}$, hence the effective degree of freedom (\textcolor{black}{i.e.} dimension of the actual phase space) is reduced. Physically, such reduction can be argued in the context of \emph{macro-hierarchy}, which is the suppression of some microscopic degree of freedom.

The nullity of $J$ implies that the vector $\dot{\bi{z}}$ has codimensions. If $\bi{z}_0 \in \mathrm{Ker}\,J$ can be \emph{integrated} as 
\begin{eqnarray}
\bi{z}_0 = \partial_{\bi{z}} C,
\label{Casimir_definition}
\end{eqnarray}
then $C\in C^\infty(M)$ is called a \emph{Casimir}. Then, the level sets of $C(\bi{z})$ foliate $M$ so that $\bi{z}(t)$ is restricted to move on only a leaf $C(\bi{z})=$ constant. In fact, 
\begin{eqnarray}
\dot{C} = \{ C, H \} = -\{ H, C \} = - \langle \partial_{\bi{z}} H, J \partial_{\bi{z}}C \rangle =0.
\label{Casimir_definition-2}
\end{eqnarray}
Notice that the invariance of a Casimir $C$ is independent of the Hamiltonian \textcolor{black}{($H$) choice}. By interpreting the degeneracy of $J$ as the suppression of some microscopic degree of freedom, the leaf of $C$ is the mathematical identification of a macro-hierarchy.

\subsection{Reduction: an example\label{subsec:reduction}}

\textcolor{black}{In this section, we invoke a simple example to explain how a Casimir} is ``created'' by a \emph{reduction}. We start with the canonical Hamiltonian system of point mass moving in $\mathbb{R}^n$. The phase space is $M=\mathbb{R}^{2n}$, and the state vector is $\bi{z}=(\bi{q},\bi{p})^{\mathrm{T}}$ with position $\bi{q}$ and momentum $\bi{p}$. On $C^\infty(M)$, we define the canonical Poisson bracket
{
\begin{eqnarray}
[G, H ] &=& \langle \partial_{\bi z} G,J_c \partial_{\bi z} H \rangle
\nonumber \\
&=& \sum_{j=1}^n (\partial_{q^j} G) \, (\partial_{{p_j}} H) 
- (\partial_{q^j} H) \, (\partial_{{p_j}} G) ,
\label{sp(2n,R)}
\\
J_c &=& \left(\begin{array}{cc}
         0&I  \\
         -I&0 
    \end{array}\right) .
\nonumber
\end{eqnarray}
Notice that the canonical Poisson operator (matrix) $J_c$ is full rank. 
\textcolor{black}{According to} the Lie-Darboux theorem, every Poisson bracket can be (at least locally) cast into the \textcolor{black}{following form:}
\begin{equation*}
[G,H]_{nc}= \langle \partial_{\bi z} G,J\partial_{\bi z} H \rangle, 
\quad J=\left(\begin{array}{cc}
          J_c&0  \\
         0&0 
    \end{array}\right) .    
\end{equation*}
In a general \textcolor{black}{non-canonical} system, the phase space $M$ may even be of an odd dimension.
If the Poisson matrix $J$ has a kernel (i.e. not full rank), such Poisson bracket is said \textcolor{black}{to be non-canonical} and the coordinates corresponding to the kernel are Casimirs.
}
Here, the canonical Poisson bracket is denoted by $[~,~]$, which will be used later in defining the Poisson bracket of the Vlasov system.

We set $n=2$ and denote the corresponding Poisson manifold by $M_4$ ($= \mathbb{R}^4$). As a trivial example of reduction, we assume that all observables are independent \textcolor{black}{of} $q^2$ and $p_2$, then the Poisson bracket \textcolor{black}{is evaluated} as
\begin{equation}
[G, H ] = 
(\partial_{q^1} G) \, (\partial_{p_1} H) - (\partial_{q^1} H) \, (\partial_{p_1} G) ,
\label{sp(2,R)}
\end{equation}
which defines a canonical Poisson algebra on the submanifold $M_2 = \{ \bi{z}_2=(q^1,p_1)^{\mathrm{T}} \} = \mathbb{R}^2$, which is embedded in $M_4$ as a leaf $\{\bi{z}\in M_4;\, q^2=c,\, p_2=c' \}$ ($c$ and $c'$ are arbitrary constants).

An interesting reduction occurs if only $q^2$ \textcolor{black}{is suppressed} in observables. The reduced phase space is the three-dimensional submanifold $M_3 = \{ \bi{z}_3=(q^1,p_1,p_2)^{\mathrm{T}} \}$. For $G$ and $H$ such that $\partial_{q^2} G = \partial_{q^2} H =0$, the Poisson bracket \textcolor{black}{is evaluated similarly to} (\ref{sp(2,R)}). The Poisson operator $J$ may be written as
\[
J = \left( \begin{array}{ccc}
0 & 1 & 0 \\
-1& 0 & 0 \\
0 & 0 & 0 \end{array} \right),
\]
whose rank is two. Therefore, $M_3$ is a degenerate Poisson manifold. The kernel of this $J$ includes the vector $(0,0,1)^{\mathrm{T}}$, which can be integrated to define a Casimir $C= p_2$. Therefore, the effective degree of freedom is \textcolor{black}{further reduced to 2. Hence,} the state vector $\bi{z}$ can move only on the two-dimensional leaf $M_2$. Evidently, the ``freezing'' of $C=p_2$ is due to the absence of its conjugate variable $q^2$.

When we observe $M_3$ from $M_4$, the reduction (i.e., the suppression of the parameter $q^2$ in the observables) means the \emph{symmetry} $\partial_{q^2}=0$. As the usual manifestation of the integral of motion, $p_2$ in $M_4$ becomes invariant if the Hamiltonian has the symmetry $\partial_{q^2} H=0$.

The conjugate variable $q^2$ corresponding to the Casimir $C=p_2$ can be regarded as the \emph{gauge parameter}. The gauge group (denoted by \textcolor{black}{$\ad_C$})---which does not change the submanifold $M_3$ embedded in $M_4$---is generated by the adjoint action
\[
\ad_C = [\circ, C] = \partial_{q^2} ,
\]
implying that the gauge symmetry is written as $\partial_{q^2}=0$. This is evident because the state vector $\bi{z}_3=(q^1,p_1,p_2)^{\mathrm{T}} \in M_3$ is ``independent'' of $q^2$.

\subsection{Gauge symmetry generated by Casimir}
\label{subsec:Casimir-gauge_symmetry}

Here, we assume $n=3$ and consider the six-dimensional phase space $M_6$. We define the angular momentum as 
\begin{equation}
{\bi \omega} = \bi{q}\times\bi{p} .
\label{6D-omega}
\end{equation}
We consider a system where every observable is a function of ${\bi \omega}$\textcolor{black}{. The} Euler top is such an example (the Hamiltonian is $H({\bi \omega})= \sum_j \omega_j^2/(2 I_j)$, where $I_1,\, I_2,\, I_3$ are the three moments of inertia). For \textcolor{black}{this} system, the effective phase space is reduced to 
\[
M_{{\bi \omega}} = \{ {\bi \omega} =\bi{q}\times\bi{p};\, (\bi{q},\bi{p})^{\mathrm{T}} \in M \} \cong \mathbb{R}^3.
\]
Let us evaluate $[~,~]$ \textcolor{black}{in} (\ref{sp(2,R)}) for the reduced class of observables $\in C^\infty(M_{{\bi \omega}a})$. The gradient of a functional $G\in C^\infty(M_{{\bi \omega}})$ is given by (denoting by$(\bi{x},\bi{y})$ the $\mathbb{R}^3$ Euclidean inner product, and by $\tilde{f}$ the perturbation of $f$)
\[
\tilde{G} = (\partial_{\bi{q}} G, \tilde{\bi{q}} ) +
( \partial_{\bi{p}} G, \tilde{\bi{p}} )
= ( \partial_{{\bi \omega}} G, \tilde{{\bi \omega}} ) .
\]
Inserting $\tilde{{\bi \omega}} = (\tilde{\bi{q}})\times\bi{p} + \bi{q}\times(\tilde{\bi{p}})$, we find
\[
\partial_{\bi{q}} G = \bi{p}\times\partial_{{\bi \omega}} G,
\quad 
\partial_{\bi{p}} G = -\bi{q}\times\partial_{{\bi \omega}} G.
\]
Therefore,
\[
[G,H ]= 
( \partial_{{\bi \omega}} G, \partial_{{\bi \omega}} H \times {\bi \omega} ) =:[G,H]_{{\bi \omega}}.
\]
We may rewrite
\begin{eqnarray}
& &[G,H]_{{\bi \omega}} = ( \partial_{{\bi \omega}} G, J({\bi \omega}) \partial_{{\bi \omega}} H ),
\nonumber \\
& &~~~ J({\bi \omega}) := - {\bi \omega}\times \circ
= \left( \begin{array}{ccc}
0 & \omega_3 & -\omega_2 \\
-\omega_3 & 0 & \omega_1 \\
\omega_2 & -\omega_1 & 0 \end{array} \right).
\label{so(3)_J}
\end{eqnarray}
We find $\mathrm{rank}\,J({\bi \omega}) =2$ (at ${\bi \omega}=0$, $\mathrm{rank}\,J({\bi \omega}) =0$). Evidently,
\[
C = |{\bi \omega}|^2
\]
is the Casimir ($J({\bi \omega}) \partial_{{\bi \omega}} C=0$). The reduction from $\bi{z}=(\bi{q},\bi{p})^{\mathrm{T}} \in \mathbb{R}^6$ to ${\bi \omega} =\bi{q}\times\bi{p} \in \mathbb{R}^3$ yields another reduction \textcolor{black}{in the} degree of freedom due to the Casimir $C$\textcolor{black}{. The} effective degree of freedom given to ${\bi \omega} \in M_{{\bi \omega}}$ is only \textcolor{black}{2}.

The Hamiltonian flow (the adjoint action) given by $C$ generates the \emph{gauge transformation} in $M_6$:
\begin{eqnarray}
\ad_C =[\circ , C] &=& \left(\sum_{j=1}^3 \partial_{p_j} C \partial_{q^j} - \partial_{q^j} C \partial_{p_j}\right)
\nonumber \\
&=& {\bi \omega}\times\bi{q} \cdot\partial_{\bi{q}}+ {\bi \omega}\times\bi{p}\cdot\partial_{\bi{p}}.
\label{gauge_symmetry_of_omega}
\end{eqnarray}
\textcolor{black}{We can demonstrate that} $[\omega_j, C]=0$ ($j=1,2,3$) \textcolor{black}{through direct calculations}.
This gauge transformation has the following geometrical meaning\textcolor{black}{. Following} (\ref{gauge_symmetry_of_omega}), the transformation $\bi{z} \mapsto \bi{z}+\epsilon \tilde{\bi{z}}$ ($\tilde{z}_j=[ z_j, C]$) gives a co-rotation of $\bi{q}$ and $\bi{p}$ around the axis ${\bi \omega}$ (note that this rotation is in the space $M_6$, not \textcolor{black}{$M_{{\bi \omega}}$}),
hence ${\bi \omega}=\bi{q}\times\bi{p}$ does not change. The rotation angle can be written as
\[
\theta = \frac{1}{2|{\bi \omega}|} \tan^{-1} \left( \frac{ ( {\bi \omega}\times\bi{q})_j}{ q_j |{\bi \omega}| } \right) 
\]
(we choose the coordinate $q_j\neq0$). 
We find $[\theta, C] =1$. \textcolor{black}{Consider embedding} $M_{{\bi \omega}}$ into a four-dimensional space 
\[
\widetilde{M_{{\bi \omega}}} 
= \{ ({\bi \omega},\theta) ;\, {\bi \omega} \in M_{{\bi \omega}},\, \theta\in {[ 0,2\pi |)} \}.
\]
For $G({\bi \omega},\theta) \in C^\infty(\widetilde{M_{{\bi \omega}}})$, we obtain
\[
[G , C ] = \sum_{j=1}^3 \partial_{\omega_j} G [ \omega_j, C] + \partial_\theta G [ \theta, C]
= \partial_\theta G .
\]
Therefore, the gauge symmetry $[\circ, C]$ can be rewritten as $\partial_\theta =0$. Reversing the perspective, for every Hamiltonian $H({\bi \omega},\theta) \in C^\infty(\widetilde{M_{{\bi \omega}}})$ that has the gauge symmetry $\partial_\theta H=0$, $C$ is invariant:
\[
\dot{C} = [C, H] = -\partial_\theta H =0.
\]

{
\textcolor{black}{This example shows} that the conjugate variable of a Casimir in the inflated phase space represents the gauge symmetry of the reduced variables, and the symmetry breaking yields a change in the Casimir.
}

{
A general condition is not known for a map (immersion) to be a \emph{reduction}.  When \textcolor{black}{making} a new binary operator from the Poisson bracket of the original Poisson manifold by variable transformation to local coordinates of the submanifold, the Jacobi identity does not necessarily hold.  Only some (physically important) examples are known as successful reductions.\cite{marsden1974reduction}  As we have seen in Sec.\ref{subsec:reduction}, a possible strategy of constructing a reduction is \textcolor{black}{as follows. First}, apply a canonical transformation of variables, \textcolor{black}{then} suppress one canonical variable (the remaining phase space has an odd dimension)\textcolor{black}{. Thus,} the conjugate variable becomes a Casimir, and the level-set of the Casimir becomes a symplectic leaf.}

\section{Water-bag model of one-dimensional beam propagation}

The aim of this work is elucidate the hierarchical structure encompassing the microscopic (kinetic) Vlasov model and the macroscopic fluid model. \textcolor{black}{For this, we} use the \emph{water-bag model} of one-dimensional charged-particle beam propagation (applicable for a nonlinear electron plasma\cite{wb1} or the drift-kinetic model of plasma\cite{wb2}). We start by reviewing the Hamiltonian formalism of the Vlasov system.

\subsection{Vlasov system as an infinite-dimensional Poisson manifold
\label{Vlasov system as an infinite-dimensional Poisson manifold}}

Let $\bi{z}=(\bi{x}, \bi{v})=(x^1,\cdots,x^n,v_1,\cdots,v_n)$ denote the coordinates of $M=X\times V=\mathbb{T}^n\times\mathbb{R}^n$, the phase space of a particle. Assuming the non-relativistic limit, the particle mass is normalized to 1, so the velocity $\bi{v}$ \textcolor{black}{and} the momentum \textcolor{black}{are parallel}. We call a real-valued function $\psi(\bi{z})\in C^\infty(M)$ an observable.
We endow the space $C^\infty(M)$ with the canonical Poisson bracket
\begin{equation}
[\psi, \varphi] = \sum_{j=1}^n (\partial_{x^j} \psi) \, (\partial_{v_j} \varphi) 
- (\partial_{v_j} \psi) \, (\partial_{x^j} \varphi) 
\label{kinetic_Lie-bracket}
\end{equation}
and denote $\mathfrak{g} = C^\infty(M)$; cf.\ (\ref{sp(2n,R)}).
The dual space $\mathfrak{g}^*$ is the set of \emph{distribution functions}; for an observable $\psi\in \mathfrak{g}$ and a distribution function $f\in\mathfrak{g}^*$, 
\begin{equation}
\langle \psi, f \rangle = \int_M \psi(z) f(z)\, \rmd z
\label{kinetic_jnner-product}
\end{equation}
evaluates the mean value of $\psi$ over the distribution function $f$.

On the space $\mathfrak{V}=C^\infty(\mathfrak{g}^*)$ of functionals,
we define 
\begin{equation}
\{ G,H \} = \langle [\partial_f G, \partial_f H], f \rangle ,
\label{kinetic_Lie-Poisson}
\end{equation}
where $\partial_f H \in T^*{\mathfrak{V}} =\mathfrak{g}$ is the gradient of $H\in \mathfrak{V}$. The bracket $\{~,~\}$ satisfies the conditions required for a Poisson bracket, hence $\mathfrak{g}^*$ is a Poisson manifold (infinite dimension). Integrating by parts, we may rewrite (\ref{kinetic_Lie-Poisson}) as
\begin{equation}
\{ G,H \} = \langle \partial_f G, [\partial_f H, f]^* \rangle = \langle \partial_f G, J(f) \partial_f H \rangle,
\label{kinetic_Lie-Poisson-2}
\end{equation}
where $[~,~]^*:\,\mathfrak{g}\times\mathfrak{g}^*\rightarrow\mathfrak{g}^*$ evaluates formally as $[a,b]^*=[a,b]$.

{
Comparing 
\[
\dot G =\langle \partial_f G, \dot{f} \rangle
\]
and 
\[
\dot G =\langle \partial_f G,[\partial_f H,f]^* \rangle
\]
($\forall G \in \mathfrak{V}$), we obtain the Vlasov equation 
\begin{equation}
\dot{f} = [\partial_f H, f]^*, 
\label{Vlasov-0}
\end{equation}
which describes the reaction of the distribution function $f(\bi z)$ to the motion of particles dictated by the {\textit {mean-field Hamiltonian}} $h=\partial_f H$.
}

Every $C(f) = \int_M g(f)\, \rmd z$ ($g$ is a smooth function: $\mathbb{R}\rightarrow\mathbb{R}$) is a Casimir. 
In fact, inserting $\partial_f C = g'(f)$, we find, for every $H$,
\[
\{ C, H \} = -\langle \partial_f H, [\partial_f C, f]^* \rangle = -\langle \partial_f H, [g'(f), f]^* \rangle =0.
\]

\subsection{One-dimensional Vlasov--Poisson system}

We have to incorporate a dynamical electromagnetic field coupled with the dynamics of charged particles. Neglecting the magnetic field, we consider a simple one-dimensional system in which the longitudinal electric field $E$ accelerates particles in the direction $x \in X=\mathbb{T} =\mathbb{R}/\mathbb{Z}$. By $\nabla\cdot\bi{E} = \rho/\epsilon_0$\textcolor{black}{, where $\rho$ is the charge density and $\epsilon_0$ is the vacuum permittivity). We} can relate $E(x,t)$ and $f(x,v,t)$ \textcolor{black}{by}
\begin{equation}
\partial_x E= \frac{q}{\epsilon_0} \int_V f(x,v,t) \rmd v,
\label{charge_density}
\end{equation}
where $q$ is the charge of the particle. Putting $E = -\partial_x \phi$ with the scalar potential $\phi(x,t)$, we obtain the Poisson equation
\begin{equation}
-\partial_x^2 \phi = \frac{q}{\epsilon_0} \int_V f(x,v,t) \rmd v.
\label{Poisson_eq}
\end{equation}
With the periodic boundary conditions $\phi(0)=\phi(1)$ and $\partial_x\phi(0)=\partial_x\phi(1)$, we can solve (\ref{Poisson_eq}) as
\begin{equation}
\phi(x,t) = \frac{q}{\epsilon_0} \mathcal{K} f(x,v,t)
\label{Poisson_eq_sol}
\end{equation}
with the integral operator
\begin{equation}
\mathcal{K} = (-\Delta)^{-1}\int_V \circ ~ \rmd v,
\label{Poisson_eq_sol-2}
\end{equation}
where $(-\Delta)^{-1}: \, L^2(X) \rightarrow H^2(X)/ \{ c \}$ is the self-adjoint operator ($\{ c \}$ is the one dimensional space of constant functions in $X$) such that
\begin{equation}
-\partial_x^2 (-\Delta)^{-1} \rho = \rho \quad (\rho \in L^2(X)).
\label{Poisson_eq_sol-3}
\end{equation}
We define the quadratic form
\[
\Phi( f) = \frac{q^2}{2\epsilon_0} \langle \mathcal{K} f, f \rangle .
\]
Using (\ref{Poisson_eq}) and (\ref{Poisson_eq_sol}), we may rewrite 
\[
\Phi(f) = \frac{1}{2} \int_X \phi \left( q\int_V f \,\rmd v \right) \rmd x = \int_X \frac{\epsilon_0 E^2}{2} \rmd x.
\]
By the symmetry $\langle \mathcal{K} f, g \rangle = \langle f , \mathcal{K} g \rangle$, we obtain
\[
\partial_f \Phi(f) = \frac{q^2}{\epsilon_0} \mathcal{K} f = q \phi.
\]
With the Hamiltonian
\begin{equation}
H(f) = \int_M \frac{ v^2}{2} f\,\rmd z + \Phi(f),
\label{Hamiltonian_Vlasov}
\end{equation}
the Vlasov equation (\ref{Vlasov-0}) reads
\begin{equation}
\dot{f} + v \partial_x f + q E \partial_v f =0.
\label{Vlasov-1}
\end{equation}

\subsection{Water-bag distribution function}

\begin{figure}[h]
\centering
\includegraphics[width=8cm]{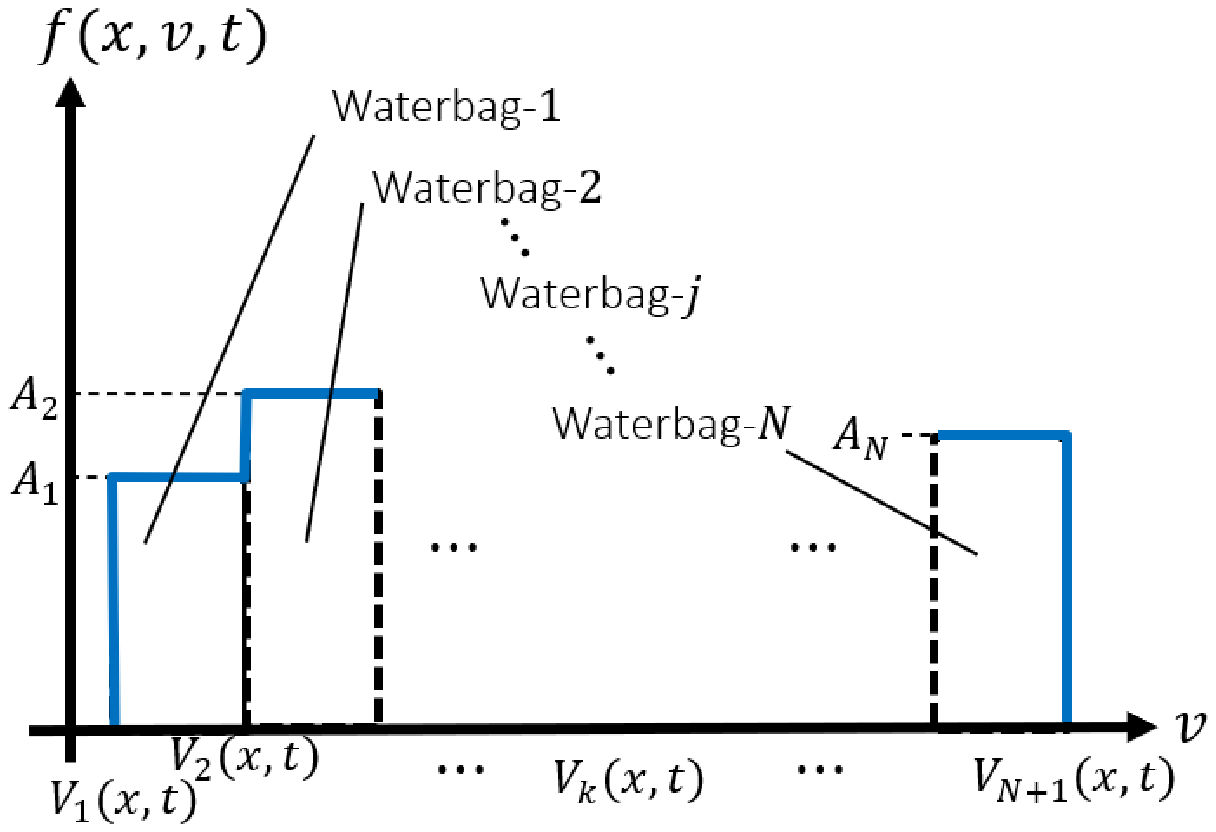}
\caption{Graph of $f(x,v)$ on a cross section of $x=$ constant. The degree of freedom of the distribution function is given by $\{V_k(x) \}$.}
\label{fig:WB}
\end{figure}

In the water-bag model (see Fig.~\ref{fig:WB}), we consider distribution functions that are linear combinations of $V$-space indicator functions: 
\begin{eqnarray}
f(x,v,t) &=& \sum_{j=1}^N A_j g_j(x.v.t) ,
\label{wbf} 
\\
g_j(x.v.t) &=& 
\left\{\begin{array}{ll}
0 & v < V_j (x,t) ,
\\
1 & V_j(x,t) \leq v \leq V_{j+1} (x,t) ,
\\
0 & V_{j+1} (x,t) < v ,
\end{array}\right.
\end{eqnarray}
where $N$ is the number of \textcolor{black}{water-bags}, and $A_j \in \mathbb{R}$ ($j=1, \cdots,N$) are constants (being amenable to Liouville's theorem). For the convenience of later calculations, we \textcolor{black}{set} $A_0= A_{N+1}=0$.
We use the index ``$j$'' to address each \textcolor{black}{water-bag}, and ``$k$'' for the boundaries; the latter runs over $1$ to $N+1$.
We define
\begin{equation}
a_k = A_{k-1} - A_{k} 
\quad (k=1, \cdots, N+1) .
\label{def_a_k}
\end{equation}
Each \textcolor{black}{water-bag} is bounded by the velocities $V_j(x,t)$ and $V_{j+1} (x,t) $ ($j=1,\cdots,N+1$), which are assumed to be smooth functions of $x$ and $t$. \textcolor{black}{We assume that each $V_j(x,t)$ is a single-valued smooth function of $x$. Therefore, the water-bag model does not allow the existence of ``islands'' in the phase space $M$. We remark that the following study on the hierarchical structure of the Vlasov-Poisson system is limited to a class of distribution functions that only have single-valued contours.} Using the step function, we may write
\[
g_j(x,v,t)=Y(v-V_j) - Y(v-V_{j+1}).
\]
For the convenience of later discussion, we fill the gap of the graph of the step function, i.e., we put $Y(0) = [0,1]$, allowing it to be multivalued.
{
The function space 
\[
\mathfrak{g}^*_N = \{ f(x,v,t) = \sum_{j=1}^N A_j g_j(x,v,t)\} 
\]
is the phase space (Poisson manifold) of the $N$-bag system. 
}

{
Let us derive the reduction of the Poisson bracket (\ref{kinetic_Lie-Poisson-2}) for observables of $N$-bag distribution functions (with fixed $A_j$). We may evaluate the perturbation of $f\in \mathfrak{g}^*_N$ as
(we \textcolor{black}{use} $\tilde{~}$ for perturbed quantities)
\begin{eqnarray*}
\tilde f &=& \sum_{j=1}^N A_j \tilde g_j \nonumber\\
&=& \sum_{j=1}^N A_j [- \delta{(v-V_j)} \; \tilde V_j + \; \delta{(v-V_{j+1})} \; \tilde V_{j+1} ]
\\
&=& \sum_{k=1}^{N+1} a_k \delta{(v-V_k)}\; \tilde V_k.
\end{eqnarray*}}
{
Since $f \in \mathfrak{g}_N^*$ is fully characterized by the \emph{contours} $V_k$ ($k=1,\cdots,N+1$), 
\[
 \mathfrak{g}_N^* \cong \{ V_k(x,t); \,k=1,\cdots,N+1 \},
\]
hence, we may put $G(f)=G(V_1,\cdots,V_{N+1})$. The chain rule reads \textcolor{black}{as}
\begin{eqnarray*}
\tilde G = \langle \partial_f G, \tilde f \rangle
&=& \langle \partial_f G, \sum_{k=1}^{N+1} a_k \; \delta{(v-V_k)} \; \tilde V_k \rangle
\\
&=& \sum_{k=1}^{N+1} ( \partial_{V_k} G, \tilde V_k ){,}
\end{eqnarray*}
where
\[
 (\bm{u},\bm{v}) = \int_X \bm{u}(x)\cdot\bm{v}(x) \rmd x.
\]
}
Therefore, denoting $\partial_{V_k} G = G_k$, we may write
\begin{equation}
\partial_f G |_{v=V_k} = \frac{1}{a_k} G_k
\quad (k=1,\cdots,N+1).
\label{chain_rule}
\end{equation}
Inserting{
\begin{eqnarray*}
\partial_v f &=& \sum_{j=1}^N A_j [\delta{(v-V_j)} -  \delta{(v-V_{j+1})} ] ,
\\
\partial_x f &=& - \sum_{j=1}^N A_j [\delta{(v-V_j)}\partial_x V_j - \delta{(v-V_{j+1})} \partial_x V_{j+1} ] ,
\end{eqnarray*}}
we obtain, for $G, H \in C^\infty(\mathfrak{g}^*_N)$,
{
\begin{eqnarray*}
\{G, H \}_N &=& \langle \partial_f G , [\partial_f H, f ]^* \rangle
\\
&=& \sum_{j=1}^N A_j
\int_X \left({\frac{1}{a_j^2}}  G_j \partial_x H_j -
\frac{1}{a_{j+1}^2} G_{j+1} \partial_x H_{j+1} \right)\,\rmd x
\\
&=& \sum_{k=1}^{N+1} -\int_X \frac{1}{a_k}
G_k \partial_x H_k \,\rmd x,
\end{eqnarray*}
}
where {(\ref{def_a_k})} \textcolor{black}{was used}. In a more illuminating form, we may write
\begin{equation}
\{ G, H \}_N = ( \nabla_{\bi{V}} G , J_N \nabla_{\bi{V}} H ),
\label{WB_bracket}
\end{equation}
where $\nabla_{\bi{V}} = (\partial_{V_1}, \cdots, \partial_{V_{N+1}})^{\mathrm{T}}$ and
\begin{equation}
J_N = \left( \begin{array}{ccc} \frac{-1}{a_1}\partial_{x} & ~ & 0
\\
~ & \ddots & ~ 
\\
0 & ~ & \frac{-1}{a_{N+1}}\partial_{x} \end{array} \right) .
\label{WB_J}
\end{equation}
{We note that the one-dimensional Vlasov system is special in that it has the hierarchical sub-algebras. The boundaries separating plateaus in the velocity space are the contour lines of the distribution function in the phase space.  The water-bag Vlasov system is equivalent to the contour dynamics system dictated by the Poisson bracket $\{~,~\}_N$ \textcolor{black}{in} (\ref{WB_bracket}) (\textcolor{black}{see} Ref.\,\citenum{sato2021contour} for the Lagrangian method to solve the contour dynamics), in which the finite degree of freedom ($N$) of the velocity space defines the hierarchy. \textcolor{black}{However in} a higher dimension, the contours of plateaus are (hyper) surfaces that can be deformed by the phase-space flow.  Therefore, the degree of freedom \textcolor{black}{of} the contours remains infinite.
While hierarchical sub-algebras are not known to higher-dimensional kinetic systems, the ideal (barotropic and inviscid) fluid model is known to be a sub-algebra of the Vlasov system.  In a three-dimensional configuration space, the ideal fluid model has the helicity as the Casimir,\cite{morrison1998hamiltonian}
which is fragile in the Vlasov system.  The gauge group generated by the helicity was studied in Ref.\,\citenum{yoshida2021helicity}.
}

\subsection{Hamiltonian}

The Hamiltonian {of the $N$-bag system} is represented \textcolor{black}{by}
{
\begin{eqnarray}
H(V_1,\cdots,V_{N+1})&=&\nonumber\\
 \frac{1}{6}\int_X &\sum_{k=1}^{N+1} & a_k V_k^3 \rmd x + \Phi(V_1,\cdots,V_{N+1}),
\end{eqnarray}
}
which is the Vlasov--Poisson Hamiltonian (\ref{Hamiltonian_Vlasov}) whose distribution function is applied to the distribution function of the water-bag model.

\subsection{Casimirs}

Evidently, 
\[
\mathrm{Ker} \, J_{N} = \{ \bi{c}=(c_1,\cdots,c_{N+1})^{\mathrm{T}};
\, c_k \in \mathbb{R} \,(k=1,\cdots,N+1) \}.
\]
This is easily integrated to derive $N+1$ independent Casimirs
\begin{equation}
\overline{V}_k (V_1,\cdots,V_{N+1})
= \int_X V_k(x,t) \rmd x 
\quad (k=1,\cdots,N+1) .
\label{Casimire_mass-2}
\end{equation}
Physically, each $\overline{V}_k$ \textcolor{black}{refers to} the average velocity (momentum) of the particles aligned along the contour (in the phase space $M$) of the distribution function $f(x,v,t)$. Because the periodic $E$ yields no net acceleration for each particle, the average velocity $\overline{V}_k$ remains constant.

Any smooth function $G( \overline{V}_1,\cdots,\overline{V}_{N+1})$ is a Casimir of the $N$-bag system. The density of water-bag $j$ is given by
\begin{eqnarray}
\rho_j = \int_{V_j}^{V_{j+1}} f\,\rmd v = A_j(V_{j+1}-V_j) \quad (j=1,\cdots,N),
\label{waterbag_density}
\end{eqnarray}
which \textcolor{black}{offers} a more convenient representation of $N$ independent Casimirs (representing the total particle number in water-bag $j$), i.e.,
\begin{eqnarray}
\overline{\rho}_j
= \int_X \rho_j \,\rmd x = A_j (\overline{V}_{j+1}-\overline{V}_j) \quad (j=1,\cdots,N).
\label{waterbag_particle_Casimir}
\end{eqnarray}


However, for the purpose of our analysis, these Casimirs $\overline{\rho}_j$ ($j=1,\cdots,N$) are not useful because they all remain constant in any higher-$N$ system (that is a sub-algebra of the Vlasov system $\mathfrak{V}$). In the \textcolor{black}{following}, we formulate the remaining one Casimir of the $N$-bag system, which ceases to be constant when we increase $N$. 

\subsection{Fragile Casimir characterizing hierarchy of sub-algebras}

{
In this section, we identify a \emph{fragile Casimir} that is not invariant as the degree of freedom $N$ of the water-bag model increases.
We define
\begin{equation}
U_j = \int_X u_j \rmd x,
\label{U_jDef}
\end{equation}
where
\[
u_j = \frac{\int_{V_j}^{V_{j+1}} v f \rmd v}{\int_{V_j}^{V_{j+1}} f \rmd v} .
\]
Physically, $U_j$ represents the average velocity of the particles contained in water-bag $j$. 
Explicitly, we may evaluate
\begin{eqnarray}
U_j&=& \frac 1 2 \int_X\frac{{A_j (V_{j+1}^2 - V_{j}^2)}}{A_j(V_{j+1} - V_j) }\rmd x
\nonumber\\
&=&\frac 1 2 \int_X(V_j + V_{j+1})\rmd x.
\end{eqnarray}
Notice that the linear transformation $\{ \overline{V}_1, \cdots, \overline{V}_{N+1} \} \mapsto \{ \overline{\rho}_1, \cdots, \overline{\rho}_{N}, U_j \}$ is a bijection of total Casimirs.
}

\begin{figure}[h]
\centering
\includegraphics[width=8cm]{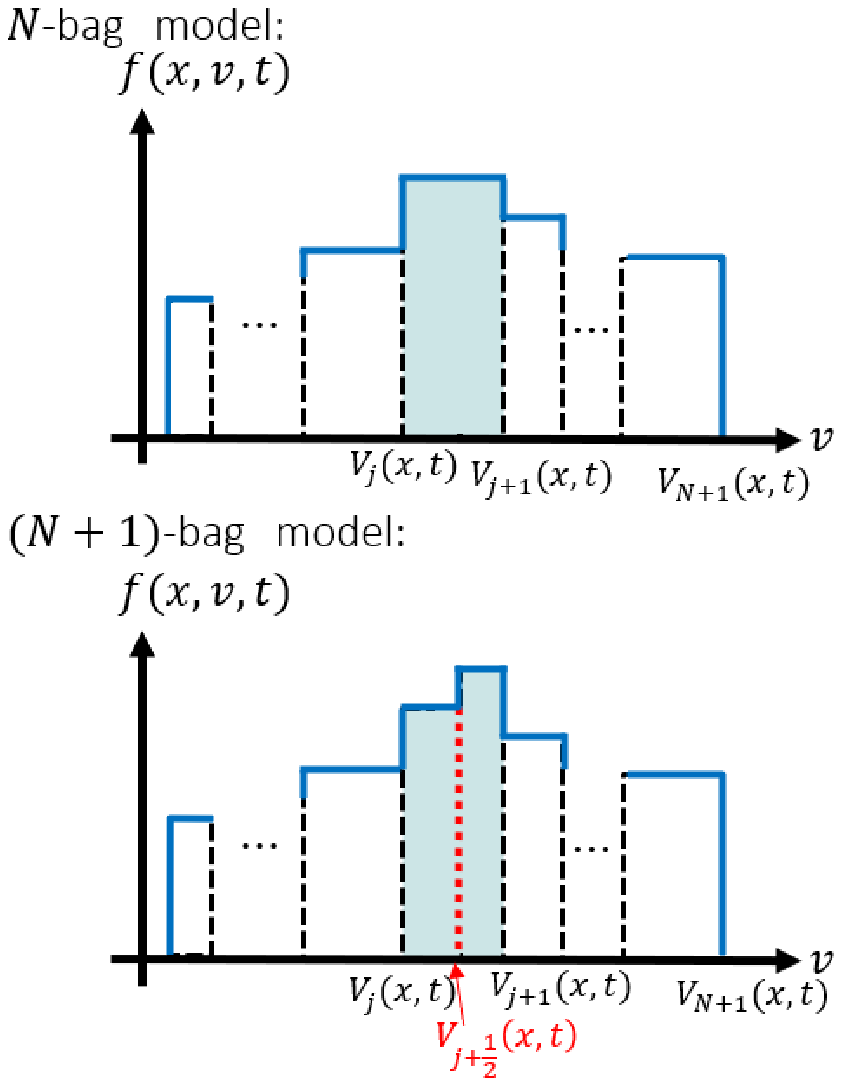}
\caption{Upper: distribution function $f$ of \textcolor{black}{an} $N$-bag system. 
Lower: adding a new boundary $,V_{j+\frac{1}{2}}$ between $V_j$ and $V_{j+1}$, we define a
distribution function $f$ of $N+1$ bags. 
In both the upper and lower figures, the light-blue region is the integration range of $U_j$ \textcolor{black}{(see definbition (\ref{U_jDef}))}.
$U_j$ is a Casimir invariant in the $N$-bag system, but not in the $N+1$-bag system.
}
\label{fig:WB_compare}
\end{figure}

{Let us introduce an additional contour $V_{j+\frac{1}{2}}(x,t)$ between $V_j(x,t)$ and $V_{j+1}(x,t)$, and divide the water-bag $j$ into two bags; see Fig.~\ref{fig:WB_compare}.  
The component $A_j g_j(x,v,t)$ of the distribution function is modified as 
\begin{equation}
A'_jg'_j (x,v,t) + A'_{j+\frac{1}{2}} g'_{j\textcolor{black}{+\frac{1}{2}}} (x,v,t) .
\end{equation} 
We \textcolor{black}{use} 
\begin{equation}
   a'_j=A'_j-A'_{j+\frac 1 2},
\end{equation}
and 
\begin{equation}
    a'_{j+ \frac 1 2}=A'_{j+\frac 1 2}-A'_{j+1}.
\end{equation} 
The previous $U_j$ of (\ref{U_jDef}) is now evaluated as
\begin{eqnarray}
U_j&=&\frac 1 2 \int_X\frac{
A'_j\left(V_{j+\frac 1 2}^2-V_j^2\right)+A'_{j+\frac 1 2}\left(V^2_{j+1}-V^2_{j+\frac 1 2}\right)
}{
A'_j\left(V_{j+\frac 1 2}-V_j\right)+A'_{j+\frac 1 2}\left(V_{j+1}-V_{j+\frac 1 2}\right)
}\rmd x\nonumber\\
&=&\frac 1 2 \int_X\frac{{- A'_j V_j^2 + a'_{j+\frac 1 2} V_{j+\frac 1 2}^2 + A'_{j+1} V_{j+1}^2}}{{- A'_j V_j + a'_{j+\frac 1 2} V_{j+\frac 1 2} + A'_{j+1} V_{j+1}}}\rmd x ,
\end{eqnarray}
which is no longer invariant for the outbreak of cross terms. 
}

{
The \emph{symmetry} required for the $(N+1)$-bag system to conserve this fragile invariant $U_j$ is written as $\{\circ,U_j\}_{N+1}=0$.
Explicitly, $\{\circ,U_j\}_{N+1}$ can be calculated as
\begin{eqnarray}
&&\{\circ,U_j\}_{N+1}=\nonumber\\&&\int_X \left(B_j \partial_{V_j} \circ +B_{j+\frac 1 2} \partial_V{_{j+\frac 1 2}} \circ +B_{j+1} \partial_{V_{j+1}} \circ\right)\rmd x ,
\label{gauge_transformation_U'}
\end{eqnarray}
where
\begin{eqnarray}
B_j&=&\frac{A'_j}{a'_j}\partial_x\frac{V_{j}-u_j}{\rho'_j + \rho'_{j+\frac 1 2}},\nonumber\\
B_{j+\frac 1 2}&=&-
\partial_x\frac{V_{j+\frac 1 2}-u_j}{\rho'_j + \rho'_{j+\frac 1 2}},\nonumber\\
B_{j+1}&=&-\frac{A'_{j+\frac 1 2}}{a'_{j+1}}\partial_x\frac{V_{j+1}-u_j}{\rho'_j + \rho'_{j+\frac 1 2}},\nonumber\\
\rho'_j&=&A'_j(V_{j+\frac 1 2}-V_j), \nonumber \\
\rho'_{j+\frac 1 2}&=&A'_{j+\frac 1 2}(V_{j+1}-V_{j+\frac 1 2}).\nonumber
\end{eqnarray}
}

{
Notice that $\{ \circ, U_j \}_{N+1}$ is not a \emph{gauge transformation} of the $N$-bag system's variables,
i.e.,  $\{ G(V_j,V_{j+1}) , U_j \}_{N+1}$ ($G \in C^\infty(\mathfrak{g}_N^*)$) is not necessarily zero
(compare with the examples of Sec.\,\ref{sec:preliminary}, where the Casimirs generated gauge transformations).
This is because the $N$-bag system is not a sub-algebra of the $(N+1)$-bag system (while both of them are \textcolor{black}{sub-algebras} of the Valsov system)
\textcolor{black}{. Recall} that water-bag $j$ of the $N$-bag system is broken into two bags in the $(N+1)$-bag system.
Therefore, the symmetry condition $\{\circ,U_ j\}_{N+1}=0$ applies not only to the new variable $V_{j+\frac{1}{2}}$, but also to $V_{j}$ and $V_{j+1}$:
\begin{eqnarray}
\partial_x \frac{u_{j}-V_{j'}}{\rho'_j + \rho'_{j+\frac 1 2}}=0 \quad (j'=j,j+\frac 1 2,j+1) .
\end{eqnarray}
We can \textcolor{black}{simplify} these relations as (for $j''=j,j+\frac 1 2,j+1,j' \neq j''$)
\begin{eqnarray}
&&\partial_x \frac{u_j-V_{j'}}{\rho'_j + \rho'_{j+\frac 1 2}} -\partial_x \frac{u_j-V_{j''}}{\rho'_j + \rho'_{j+\frac 1 2}}\nonumber \\
&=&(V_{j'}-V_{j''})\frac{\partial_x \left(\rho'_j + \rho'_{j+\frac 1 2} \right)}{\left(\rho'_j + \rho'_{j+\frac 1 2}\right)^2}-\frac{\partial_x (V_{j'}-V_{j''})}{\left(\rho'_j + \rho'_{j+\frac 1 2}\right)}\nonumber \\
&=&\frac {V_{j'}-V_{j''}} {\rho'_j + \rho'_{j+\frac 1 2}}\left(\frac{\partial_x (A'_j(V_{j+\frac 1  2}-V_j)+ A'_{j+\frac 1 2}(V_{j+1}-V_{j+\frac 1 2})) } {A'_j(V_{j+\frac 1 2}-V_j)+ A'_{j+ \frac  1 2}(V_{j+1}-V_{j+\frac 1 2}) }\right.\nonumber \\ &&\left. - \frac {\partial_x (V_{j'}-V_{j''})} {V_{j'}-V_{j''}}\right)
=0,
\label{U2_condition_process}
\end{eqnarray}
which reads
\begin{eqnarray}
&&\frac{\partial_x (A'_j(V_{j+\frac 1 2}-V_j)+ A'_{j+\frac 1 2}(V_{j+1}-V_{j+\frac 1 2})) } {A'_j(V_{j+\frac 1 2}-V_j)+ A'_{j+\frac 1 2}(V_{j+1}-V_{j+\frac 1 2}) } = \frac {\partial_x (V_{j'}-V_{{j''}})} {V_{j'}-V_{{j''}}}.\nonumber
\end{eqnarray}
Integrating with respect to $x$, we obtain
\begin{eqnarray}
&&A'_j(V_{j+\frac 1 2}-V_j)+ A'_{j+\frac 1 2}(V_{j+1}-V_{j+\frac 1 2}) \nonumber \\ && = C_{j'j''} (V_{j'}-V_{{j''}}),
\end{eqnarray}
where $C_{j'j''}$ is a constant.
The symmetry condition now reads
\begin{equation}
V_{j+\frac 1 2}=\alpha V_j + (1-\alpha) V_{j+1} \quad (\alpha = \mbox{const.},\, 0<\alpha<1),
\label{U2_condition}
\end{equation}
implying that the new contour $V_{j+\frac 1 2}$ included in the inflated $N+1$ system must be an internally dividing point of the original contours with an (arbitrary) homogeneous ratio \textcolor{black}{\[
\alpha = \frac{V_{j+1}-V_{j+\frac{1}{2}}}{V_{j+1} - V_j} .
\]} Every deformation of the contour $V_{j+1}$ from this symmetry violates the conservation of $U_j$.
}

{
\section{Symmetry breaking}
\textcolor{black}{We show that the velocity-space shear flow is a basic mechanism of symmetry breaking, \textcolor{black}{including} a change in the fragile Casimir $U_j$.
}
For an explicit demonstration, let us consider a simple model without electric field\textcolor{black}{, i.e., we consider
\[
\partial_t f + v \partial_x f = 0.
\]
The corresponding characteristic equation is the equation of motion of a free particle:
\[
\frac{\rmd}{{\rmd} t} x = v ~(=\mathrm{constant}).
\]
}
We also reset the configuration space to $\mathbb{R}$. 
We define, with a positive constant $c$,
\[
F_c (x) = \left\{ \begin{array}{ll}
c x & 0 \leq x < 1 ,
\\
c(2-\textcolor{black}{x}) & 1 \leq x < 2 ,
\\
0 & \mathrm{otherwise},
\end{array}  \right.
\]
and put
$V_c (x,0) = F_c(x)$.
With this initial condition, the contour of the solution to the field-free Vlasov equation is given by 

\[
V_c (x,t) = \left\{ \begin{array}{ll}
cx/(1+ct) & 0 \leq x < 1+ct ,
\\
c(2-x)/(1-ct) & 1+ct \leq x < 2 ,
\\
0 & \mathrm{otherwise\textcolor{black}{,}}
\end{array}  \right.
\]
\textcolor{black}{for $ 0 \leq t < 1/c$.}
We find that the ``wave breaking'' occurs at $t=1/c$.
}

\begin{figure}[h]
\centering
\includegraphics[width=8cm]{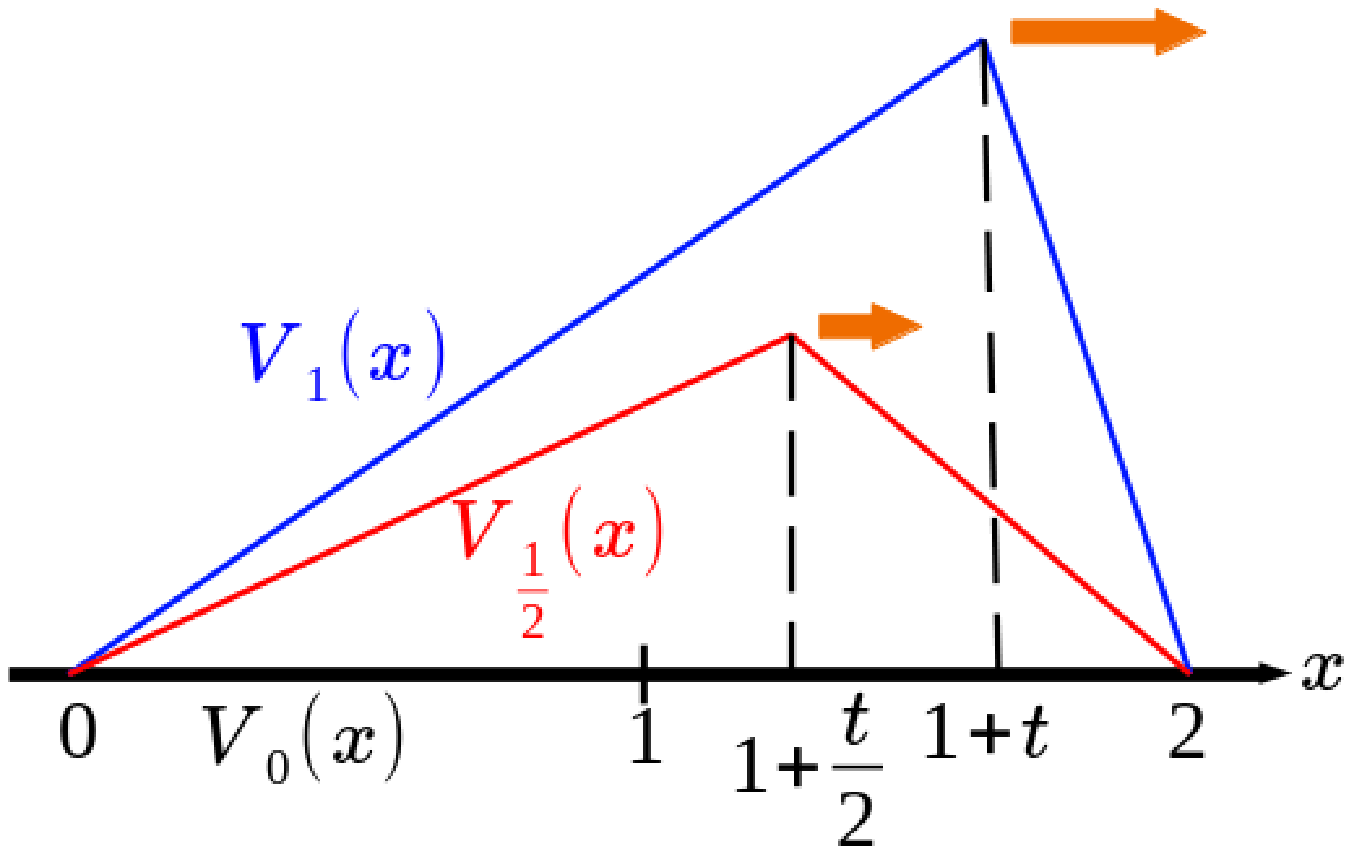}
\caption{
{
A toy model of the $2$-bag system.
Without an electric field, the contours $V_c (c=0,\frac 1 2,1)$ are moved by a linear shear flow.
}
}
\label{fig:triangle0611}
\end{figure}

\begin{figure}[h]
\centering
\includegraphics[width=8.5cm]{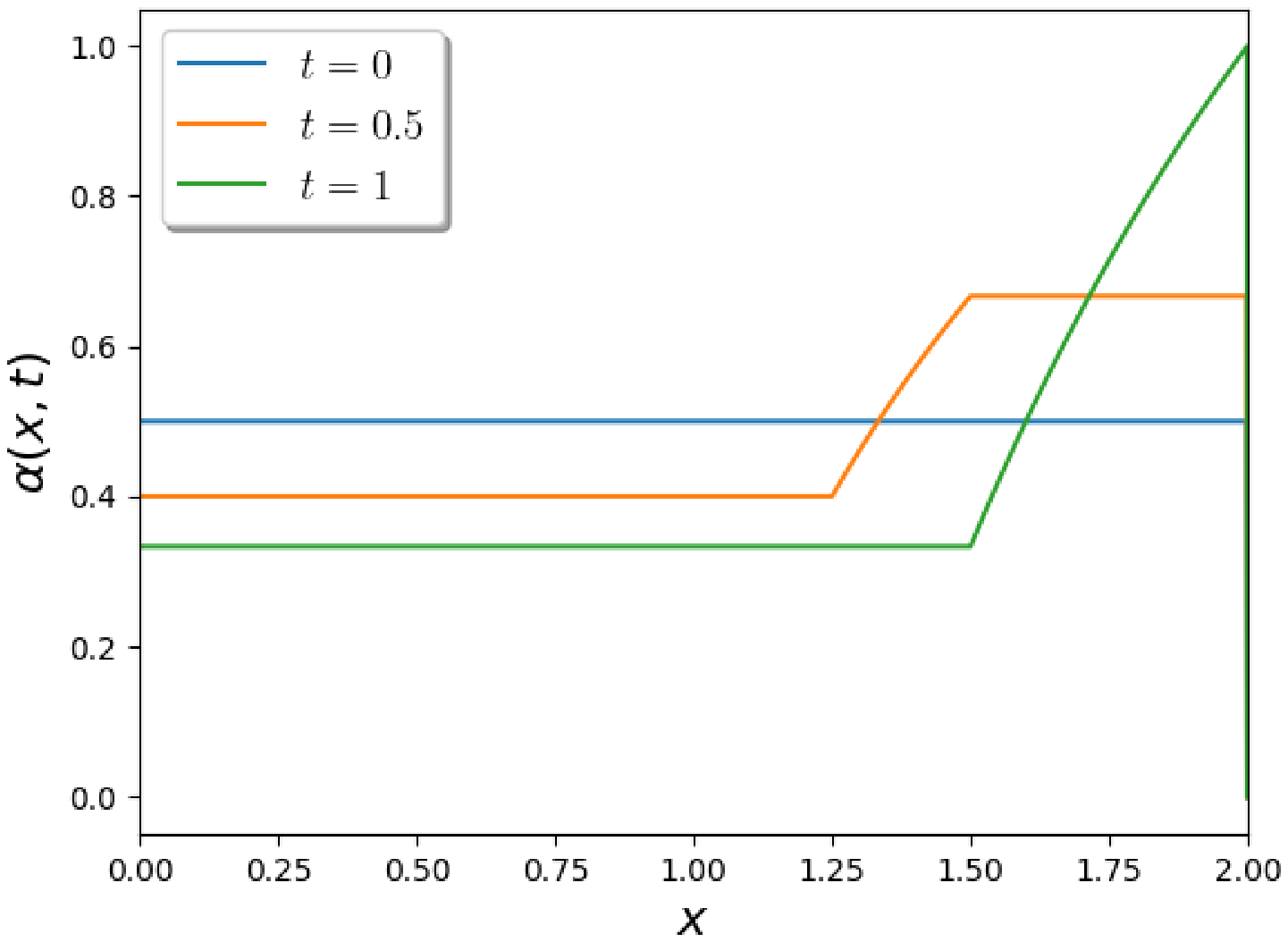}
\caption{
{\textcolor{black}{
Evolution of the split ratio $\alpha(x)$ in $t=0, 0.5, 1$.
The horizontal axis is $x$, and the vertical axis shows $\alpha$.
}}
}
\label{fig:triangleAlpha}
\end{figure}

\begin{figure}[h]
\centering
\includegraphics[width=8.5cm]{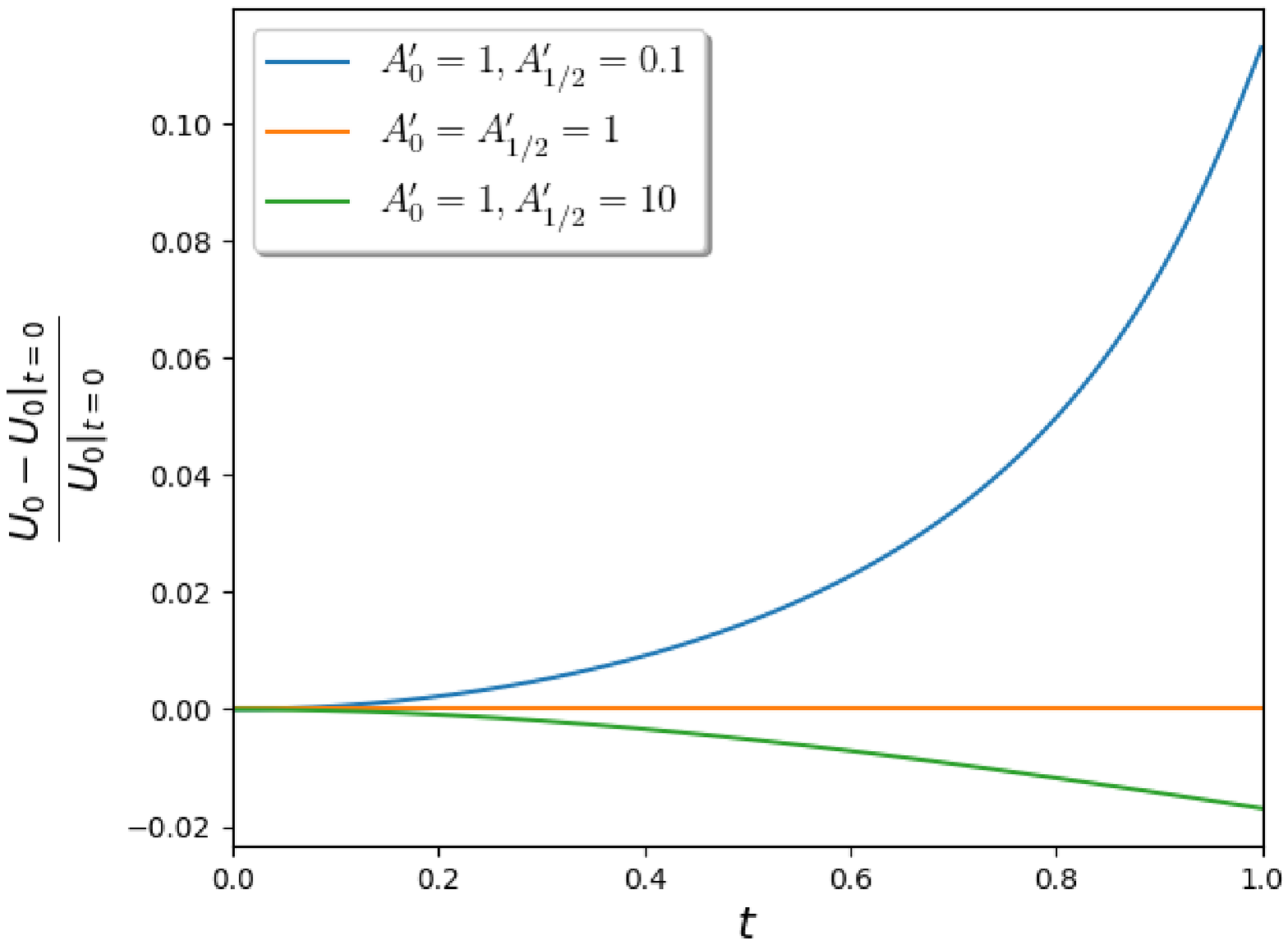}
\caption{
{
The time evolution of the fragile Casimir $U_0$ in the field-free Vlasov system.
Three different cases are compared: 
green line is for $A'_0 = A'_{1/2} =1$, purple line is for $A'_0 =1,  A'_{1/2} =0.1$, and blue line is for $A'_0 =1, A'_{1/2} =10$.
The horizontal axis is the time, and the vertical axis shows $\frac {U_0 - U_0|_{t=0}}{U_0|_{t=0}}$.
}
}
\label{fig:triangleU0611}
\end{figure}

{
We consider an $N=1$ water-bag system with $V_0(x,t)$ and $V_1(x,t)$ (notice that the index $j$ runs over 0 and 1),
and introduce an intermediate contour $V_{\frac{1}{2}}(0,t)$ to define an $N=2$ system.
In the initial condition, $U_0 = 1/2$, and $\alpha(x)\equiv 1/2$ ($0<x<2$).
As the contours move (see Fig.~\ref{fig:triangle0611}), symmetry breaking proceeds:
\[
\alpha(x,t) =
\left\{ \begin{array}{ll}
\frac{1}{2(1+t/2)} & 0 \leq x < 1+t/2 ,
\\
\frac{3x-2(1+t)}{x(2-t)} & 1+t/2 \leq x < 1+t ,
\\
\frac{1}{2(1-t/2)} & 1+t \leq x < 2 ,
\\
0 & \mathrm{otherwise} .
\end{array}  \right.
\]
\textcolor{black}{From this solution, we see how the ``velocity shear'' in the phase space proceeds the symmetry breaking, i.e., the violation of the constancy of $\alpha$ (see Fig.~\ref{fig:triangleAlpha}).} The change of $U_0$ is shown in Fig.~\ref{fig:triangleU0611}.
}

\section{Conclusion}
{
The term ``kinetic effect'' vaguely represents various phenomena stemming from the freedom in the velocity space.  In a collisional system, the kinetic effect is measured by a deviation from the Boltzmann distribution (for example, evaluating moments or cumulants).  In a collisionless system, which is independent of statistical context, a different scheme \textcolor{black}{is required} to quantify the kinetic effect.  Here we proposed a method to measure the deformation of the distribution function as a symmetry breaking.  The fragile Casimir, $U_j$ \textcolor{black}{in} (\ref{U_jDef}), is the parameter \textcolor{black}{that should} be evaluated to detect the symmetry breaking.  The relevant symmetry is the homogeneity of the contour lines in the phase space, as represented by the constancy of $\alpha$ in (\ref{U2_condition}). 
The ubiquitous shear flow in the velocity space distorts $\alpha$ and breaks the symmetry. 
}
\vspace{16pt}
\begin{acknowledgments}
The authors thank the members of the Yoshida laboratory for valuable discussions.
\end{acknowledgments}

\bibliography{2021_Waterbag_aip}

\end{document}